\documentclass[twocolumn,english,aps,prl,showpacs,superscriptaddress,floats,amsmath,amssymb,floatfix,nobalancelastpage]{revtex4-1}
\usepackage[T1]{fontenc}
\usepackage[latin9]{inputenc}
\usepackage{color}
\usepackage{amstext}
\usepackage{graphicx}
\usepackage{esint}

\begin{document}

\title{Highly Dispersive Scattering From Defects In Non-Collinear Magnets}

\author{Wolfram Brenig}
\email{w.brenig@tu-bs.de}
\affiliation{Institut f\"{u}r Theoretische Physik, Technische Universit\"{a}t
Braunschweig, 38106 Braunschweig, Germany}

\author{A. L. Chernyshev}
\affiliation{Department of Physics, University of California, Irvine, California
92697, USA}

\begin{abstract}
We demonstrate that point-like defects in non-collinear magnets give rise to a highly
dispersive structure in the magnon scattering, violating a standard paradigm of its
momentum independence.  For a single impurity spin coupled to a prototypical
non-collinear antiferromagnet, we find that the resolvent is dominated by a distinct
dispersive structure with its momentum-dependence set by the magnon dispersion and
shifted by the ordering vector. This feature is a consequence of umklapp scattering
off the impurity-induced {\it spin texture}, which arises due to the non-collinear
ground state of the host system.  Detailed results for the staggered and uniform
magnetization of this texture as well as the $T$-matrix from numerical linear
spin-wave theory are presented.
\end{abstract}


\maketitle
\emph{Introduction.}---Electron localization \cite{Anderson58}, paramagnetic
impurities in superconductors \cite{Abrikosov61}, and the orthogonality catastrophe
\cite{Anderson67}, all attest to the fundamental importance of impurities as probes
of quantum many-body systems. Major research effort in cuprate superconductors has led to
extensive studies of  impurities in the square-lattice Heisenberg
antiferromagnets (HAFs), uncovering new universality classes for disorder-driven
transitions \cite{Vajk02,Sandvik02,Vojta05,Yu05,Sandvik06}, impurity-induced magnetic
order \cite{Martins97}, fractional Curie response \cite{Sachdev99,Hoglund03}, and
anomalous low-energy magnon scattering \cite{Brenig91,Chernyshev01}.

While the square-lattice HAF is unfrustrated and has a collinear ground state,
defects in non-collinear and frustrated quantum magnets have come into focus only
recently, displaying an even richer physics. This includes frustration release, dimer
freezing, and mutual impurity repulsion \cite{Dommange03,Martins08,Mila12}, valence
bond glass states \cite{Singh10,Poilblanc10}, emergent gauge-flux pinning
\cite{Willans11}, breakdown of linear response \cite{Wollny12}, fractional impurity
moments, and --- the primary topic of this Letter --- spin textures
\cite{Henley01,Eggert07,Wollny11}.

\begin{figure}[b]
\includegraphics[width=0.9\columnwidth]{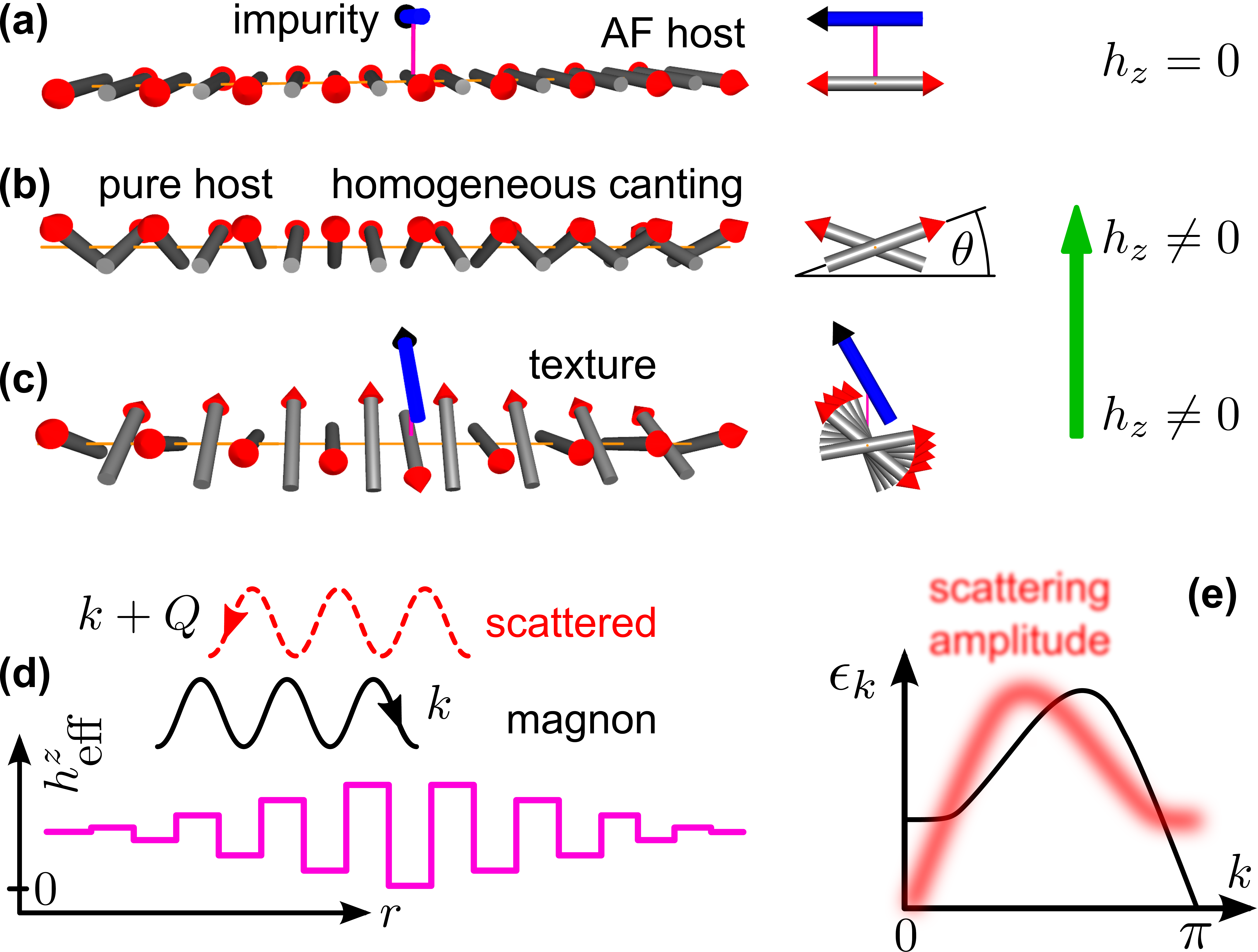}
\caption{\label{fig1}(color online)
(a) Impurity spin coupled to a collinear state: all spins co-aligned.
(b) Homogeneous canted state   in external field $h_{z}$.
(c) Impurity spin coupled to the canted state: host spins readjust, creating a
texture.
(d) 1D sketch of umklapp scattering by the texture, which generates staggered
$z$-component of the effective field with the wave vector $Q=\pi$.
(e) Solid black line: magnon dispersion; blurred red line: dispersive peak in
scattering amplitude.
}
\end{figure}

Impurity-induced spin textures are a genuine hallmark of non-collinear magnetic
order and can be understood on a purely classical level.  Removing a spin from
the host, or adding an extra defect spin, locally perturbs the balance of
exchange fields and requires the surrounding spins of the non-collinear host to
readjust their directions recursively, resulting in a long-ranged modification
of the canting angles, i.e., a texture \cite{Henley01,Eggert07,Wollny11}. A 1D
sketch of this is shown in Figs.~\ref{fig1}(b) and (c) for the field-induced
non-collinear state coupled to an impurity spin. The readjustment effect is
absent for collinear order, where impurity spin simply co-aligns with the host,
as in Fig.~\ref{fig1}(a). In contrast to that, the texture implies a fractional
screening of the impurity moment \cite{Wollny11}. The real-space decay of the
texture depends on the nature of the non-collinear state. In a field-induced
canted states, textures decay exponentially on a length scale inversely
proportional to the external field \cite{Eggert07}.  In frustration-induced
non-collinear states, Goldstone modes lead to an algebraic decay of the texture
\cite{Wollny11,Henley01,Sushkov05}.

In this Letter we advance the field beyond previous studies, which have focused on
the static properties of defects, and investigate magnon impurity-scattering in
non-collinear magnets. To be specific, we consider the field-induced canted state of
the square-lattice HAF with an additional defect, namely an extra out-of-plane spin
interacting by an exchange coupling with one of the host spins. We discover a
phenomenon rather surprising, if confronted with conventional expectations for the
scattering amplitude from a point defect, which is either momentum-independent
altogether, aside from the trivial transformation of the excitation basis, or
contains only a broad momentum modulation due superposition of a few partial waves.
Instead, the scattering amplitude displays a strongly dispersive feature, clearly
tracing the magnon dispersion shifted by the magnetic ordering vector.  We show that
this effect is an unequivocal consequence of the spin texture. Intuitively, an
effective staggering of the magnetic field is generated by the texture, made explicit
in Fig.~\ref{fig1}(d). This serves as a potential for umklapp scattering of magnons,
which, in turn, leads to the central new feature in the $T$-matrix --- a momentum-dependent
resonance.  In the following, we provide the detailed arguments for this result,
which should remain valid for a wide class of frustrated non-collinear systems, and
suggest experiments to test this prediction.

\emph{Model.}---We consider the square-lattice HAF at $T\!=\!0$ in an external
field, coupled to an impurity spin ${\bf S}^\prime$
\begin{equation}
{\cal H}=J_0\sum_{\langle lm\rangle}\mathbf{S}_{l}\cdot\mathbf{S}_{m}-h\sum_{l} S^z_{l}
+J\mathbf{S}_{0}\cdot\mathbf{S}^\prime_i -h\, S^{\prime z}_i\,,
\label{eq1}
\end{equation}
where $\langle lm\rangle$ are the nearest-neighbor bonds of the square lattice,
the exchange couplings of the host ($J_0$) and
host-to-impurity ($J$) are antiferromagnetic. The gyromagnetic ratio is identical for
all spins and is included into the magnetic field $h$. In the following, we set $J_0=1$.

The spin configuration that minimizes the classical energy of model (\ref{eq1}) at
$h{\neq}0$ corresponds to an inhomogeneous distribution of spin tilt angles $\theta_l$
out of the $xy$-plane where ordering occurs at $h\!=\! 0$, see Fig.~\ref{fig1}.  For a
$1/S$ expansion, we align the local spin quantization axis on each site in
the direction given by the local canted frame \cite{Zhitomirsky98,Mourigal10}.  The
rotation of spin components from the laboratory frame $(x_0,y_0,z_0)$ is given by
$S_i^{y_0}\!=\!S_i^{y}$ and
$S_l^{x_0(z_0)}\!=\! S_l^{x(z)} \sin\theta_l\! \pm\! S_l^{z(x)} e^{i{\bf Q}\cdot {\bf r}_l}
\cos\theta_l$,
where ${\bf Q}\!=\!(\pi,\pi)$ is the N\'eel ordering wave-vector.
The transformation is the same for the impurity spin
${\bf S}^\prime_i$ as it can be seen as a neighbor of the site $l\!=\!0$, which it is
coupled to.

Expressing the spin operators in terms of Holstein-Primakoff bosons,
Hamiltonian (\ref{eq1}) is transformed into a series ${\cal H}\!=\!{\cal
  H}_{\mathrm{class}}+{\cal H}_{1}+{\cal H}_{2}+...$ with decreasing powers of $S
(S^\prime)$ and increasing number of boson operators. Each term in
this series depends on all $\theta_{\{l\}}$ and
${\cal H}_{\mathrm{class}}$ is the classical energy \cite{supp}. The harmonic spin-wave term is
${\cal H}_{2}$ and  stability requires  ${\cal H}_{1}{\equiv} 0$.
Equivalently,  the ground state  must minimize
${\cal H}_{\mathrm{class}}$, i.e. $\partial {\cal H}_{\mathrm{class}} / \partial
\theta_{\{l\}}=0$.  Without the impurity, all $\theta_l\equiv\sin^{-1}(h/h_s)$ with
the saturation field $h_s=8 S$ \cite{Zhitomirsky98}. With the impurity, minimization  gives a set
of nonlinear coupled equations, which determine the inhomogeneous distribution of the
local tilt angles $\theta_{l}$ --- referred to as the \emph{texture} hereafter.

In what follows, we study the properties of this texture numerically in finite
$N\times N$ clusters with
periodic boundary conditions. First, we briefly address its static properties and then
turn to its quantum dynamics using numerical real-space diagonalization of ${\cal
  H}_{2}$.

\begin{figure}[t]
\includegraphics[width=.98\columnwidth]{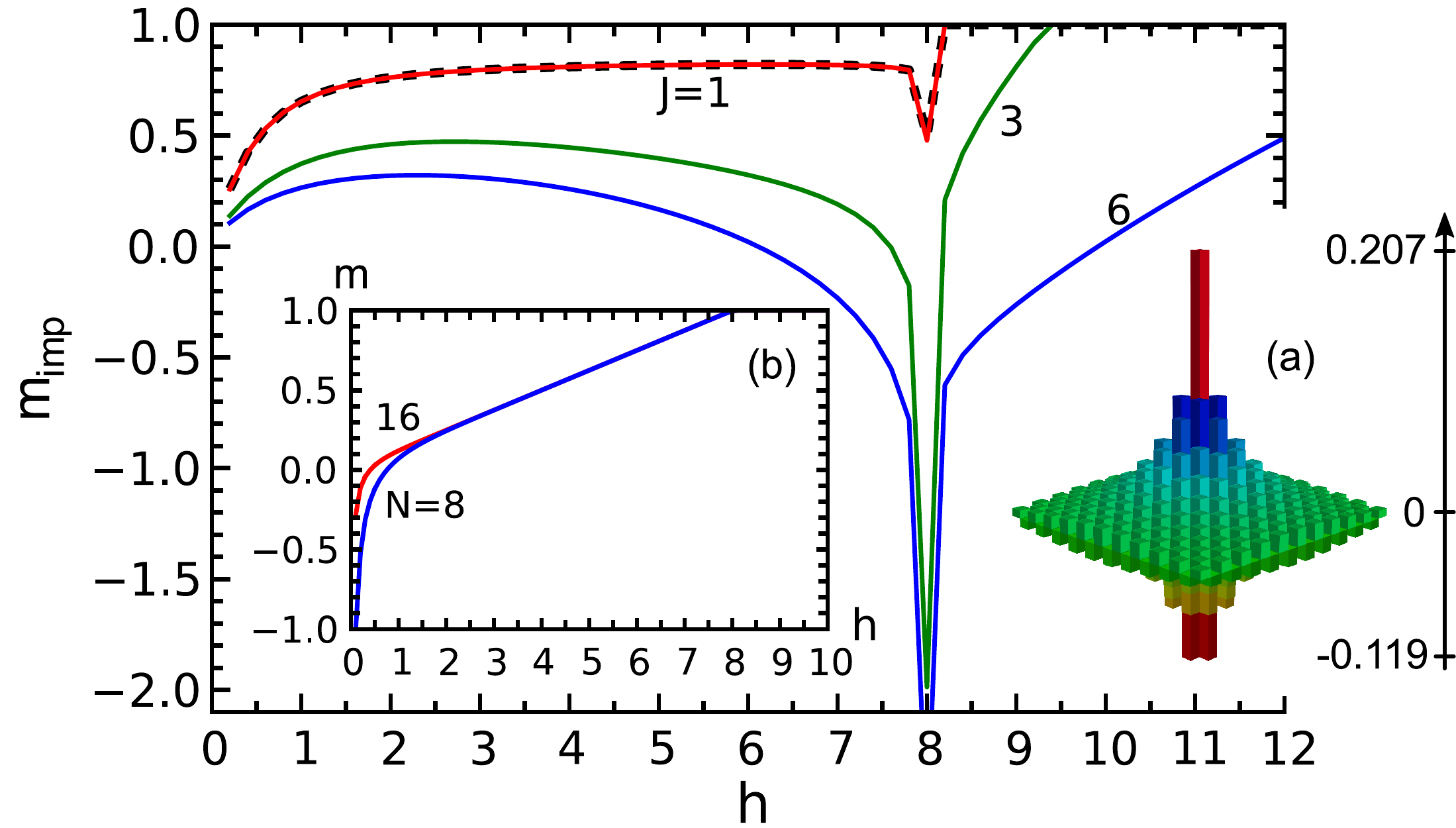}
\caption{\label{fig2}(color online)  Impurity magnetization $m_{imp}$ vs $h$ for $J$=1, 3, and 6
in $N\!=\!72$ cluster and for $J$=1 in $N\!=\!64$ cluster (dashed). Insets: (a) Local magnetization
$\Delta m_l^z\!=\!(\sin(\theta_0)\!-\!\sin(\theta_l))$ in a $21{\times}21$ section of the $N\!=\!72$ cluster, for $J$=1, $h$=0.4.
(b) Local magnetization $m_l^z$ at the distance ($N/2, N/2$) from $l\!=\!0$ in $N\!=\!8$ and 16 clusters
for $J$=1.}
\vskip -0.2cm
\end{figure}

\emph{Classical texture.}---The spatial extent and field-dependence of the texture can
be described in terms of the {\em staggered} $z$-component of the magnetization
$m_{stag,{\bf r}_l}^{z}$ obtained from the set of $\sin(\theta_l)$.  Our results, inset (a) of Fig. \ref{fig2}
 and \cite{supp}, largely corroborate earlier findings of \cite{Eggert07},
where $m_{stag,{\bf r}_l}^{z}$ was investigated by a continuum theory and quantum Monte
Carlo for a different impurity type. In particular,
the texture decays exponentially at $|{\bf r}_l|{\gg}1$, consistent with the impurity
not coupled to the Goldstone mode of the host system.

Fig.~\ref{fig2} shows another characteristics of the texture: the impurity
contribution to the \emph{uniform} magnetization
$m_{imp}{=}m^{z}{-}m_{host}^{z}$ vs field for several values of the coupling
$J$. Here $m^{z} {=} \sum_{l}S_{l}^{z}$ is the uniform magnetization including
$S'^z_i$ and $m_{host}^{z} {=} \sum_{l\neq i}S_{l}^{z}$ is that of the host in
the absence of impurity.  We use $S {=} S^\prime {=} 1$ hereafter.  Defining the
impurity susceptibility as $\chi_{imp}\!=\!\partial m_{imp}/\partial h$,
Fig.~\ref{fig2} shows several regimes of screening of the impurity by the
texture: partial, complete, and overscreening, as evidenced by
$\chi_{imp}\!>\!0$, $\approx\! 0$, and $<\!0$, respectively. This is consistent
with a field-dependent fractional effective impurity spin \cite{Wollny11}, and
is in a stark contrast with the collinear HAFs, where classical
$m_{imp}\!\equiv\! S^\prime$.  The impurity magnetization is critical at $h_s$
because the susceptibility of the host diverges as ${\sim}\ln|h-h_s|$
\cite{Zhitomirsky98}. Fig.~\ref{fig2} also shows that the saturation in the
system with impurity occurs above $h_s$ of the pure host and that
finite-size effects are negligible for the clusters and field ranges that we use.

For completeness, we note that the impurity-induced classical texture behaves
singularly at $h\!\rightarrow\! 0$, although in a field range of measure zero in
the thermodynamic limit --- an effect also noted in \cite{Wollny12,weaware}. In
a finite system, the energy gain of the canted state in Fig.~\ref{fig1}, $\Delta
E\! \sim \!-N^{2}h^{2}/(8S)$, is less than that of the state in which the N\'eel
order of the host and the impurity spin both fully align with the field, $\Delta
E\! =\! -hS^\prime$. Thus, at $h\!=\!0^+$ host spins are aligned (anti-aligned)
with the field, $S_l^z\!=\!\pm S$.  A spin-flop crossover to the textured state
occurs at $h_{c} \!\sim\! 8SS'/N^{2}\!\rightarrow\!0$ as
$N\!\rightarrow\!\infty$.  Inset (b) of Fig.~\ref{fig2} displays this behavior
on \emph{judiciously small} systems by monitoring the magnetization $m_l^{z}$ of
a spin at the largest geometrical distance from the impurity.

\emph{$T$-matrix.}---We now turn to the spectral properties of the
system. Because the texture breaks translational invariance, the Bogolyubov
transformation of ${\cal H}_2$ has to be performed numerically
\cite{Colpa78}. The \emph{para-unitary}, $2(N^{2}\!+\!1){\times} 2(N^{2}\!+\!1)$
matrix $\mathbf{U}$ of this transformation maps the local Holstein-Primakoff
bosons ${\bf a}^{\dagger} = [a_{1}^{\dagger}, \dots,a_{N^{2}}^{\dagger},
a_{i}^{\dagger}, a_{1}^{\phantom{\dagger}},
\dots,a_{N^{2}}^{\phantom{\dagger}},a_{i}^{\phantom{\dagger}}]$ onto Bogolyubov
bosons $\mathbf{\bar{b}}^{\dagger} = [\bar{b}_{1}^{\dagger}, \dots,
\bar{b}_{N^{2}+1}^{\dagger}, \bar{b}_{1}^{\phantom{\dagger}}, \dots,
\bar{b}_{N^{2}+1}^{\phantom{\dagger}}]$, whose Hamiltonian, ${\cal H} =
\mathbf{\bar{b}}^{\dagger}\mathbf{E}\mathbf{\bar{b}}/2$, is diagonal.  The
eigenenergies $E_n$ are all positive except for $E_{j}\!=\!0$ of the Goldstone
mode. The Green's function in the $\mathbf{\bar{b}}$-basis is also a diagonal
$2(N^{2}\!+\!1){\times}2(N^{2}\!+\!1)$ matrix
$\mathbf{G}^{\mathbf{\bar{b}}}(z)=[z\mathbf{P}-\mathbf{E}]^{-1}$, where
$\mathbf{P}$ is the para-unit matrix with $1(-1)$ in the upper (lower) half of
its diagonal. The Green's function of the local Holstein-Primakoff bosons is
$\mathbf{G}^{\mathbf{a}}(z)=(\mathbf{U}^{\dagger})^{-1}\mathbf{G}^{\mathbf{\bar{b}}}(z)
\mathbf{U}^{-1}$.

However, to formulate the scattering problem for the impurity-induced texture, the
proper basis is that of the Bogolyubov magnons of the \emph{uniform} host, which
describe the incident and scattered magnons as plane-wave eigenstates of momentum $\mathbf{k}$.
Thus, we first Fourier transform the matrix elements of
$\mathbf{G}^{\mathbf{a}}(z)$ of the local {\em host}
bosons to $\mathbf{k}$-space, yielding a matrix
$\mathbf{G}^{\mathbf{a}}_{\mathbf{k}'\mathbf{k}}(z)$. Second, the host boson terms of this matrix are mapped
onto the basis of the Bogolyubov  magnons
${\bf b}^{\dagger}=[b_{\mathbf{k}}^{\dagger}, b_{\mathbf{k}}^{\phantom\dagger}]$ of the {\em uniform} host,
using the known  parameters of the transformation,  $u_{\bf k}$ and $v_{\bf k}$, for the
square-lattice HAF in a field \cite{Mourigal10,supp}. This yields a matrix Green's
function with three $2\times2$ substructures made from blocks of rank $N^{2}\times
N^{2}$, $1$, and $N$. They correspond to the {\em dressed} (i) host magnon, (ii)
impurity, (iii) and magnon-impurity Green's functions
$\mathbf{G}_{\mathbf{k}'\mathbf{k}}(z)$, $\mathbf{G}_i(z)$, and
$\mathbf{G}_{\mathbf{k}i}(z)$, respectively.

Altogether, starting from the numerical solution of the classical texture, followed by the real-space
diagonalization of the harmonic problem, and Bogolyubov
transformation onto the uniform host, we obtain the dressed magnon Green's function
$\mathbf{G}_{\mathbf{k} '\mathbf{k}}(z)$.  On the other hand, $\mathbf{G}_{\mathbf{k}
'\mathbf{k}}(z)$ can be written in the conventional form
\begin{eqnarray}
\mathbf{G}_{\mathbf{k}'\mathbf{k}}(z) =
\delta_{\mathbf{k}'\mathbf{k}}\mathbf{G}_{\mathbf{k}}^{0}(z)+
\mathbf{G}_{\mathbf{k}'}^{0}(z)\mathbf{T}_{\mathbf{k}'\mathbf{k}}(z)\mathbf{G}_{\mathbf{k}}^{0}(z)\, ,
\label{eqGkk}
\end{eqnarray}
where $\mathbf{G}_{\mathbf{k}}^{0}(z)$ is the diagonal $2\times2$ Green's function of
the uniform host magnons with
$G_{\mathbf{k}}^{0,11}(z)=G_{\mathbf{k}}^{0,22}(-z)=
[z-\varepsilon_{\mathbf{k}}]^{-1}$  and $\varepsilon_{\mathbf{k}}$
is the magnon energy.
Using Eq.~(\ref{eqGkk}), we can now extract the  scattering matrix
$\mathbf{T}_{\mathbf{k}'\mathbf{k}}(z)$ from $\mathbf{G}_{\mathbf{k}'\mathbf{k}}(z)$.  For the remainder
of this work we focus on the diagonal elements of the $T$-matrix,
$\mathbf{T}_{\mathbf{k\mathbf{k}}}(z)= \mathbf{T}_{\mathbf{k}' \mathbf{k}}(z)
\delta_{\mathbf{k}' \mathbf{k}}$
\begin{equation}
\mathbf{T}_{\mathbf{k}\mathbf{k}}(z)=[\mathbf{G}_{\mathbf{k}}^{0}(z)]^{-2}
\mathbf{G}_{\mathbf{k}\mathbf{k}}(z)-[\mathbf{G}_{\mathbf{k}}^{0}(z)]^{-1}\,,
\label{eq6}
\end{equation}
which suffice to state our main findings.

\begin{figure}
\includegraphics[width=0.9\columnwidth]{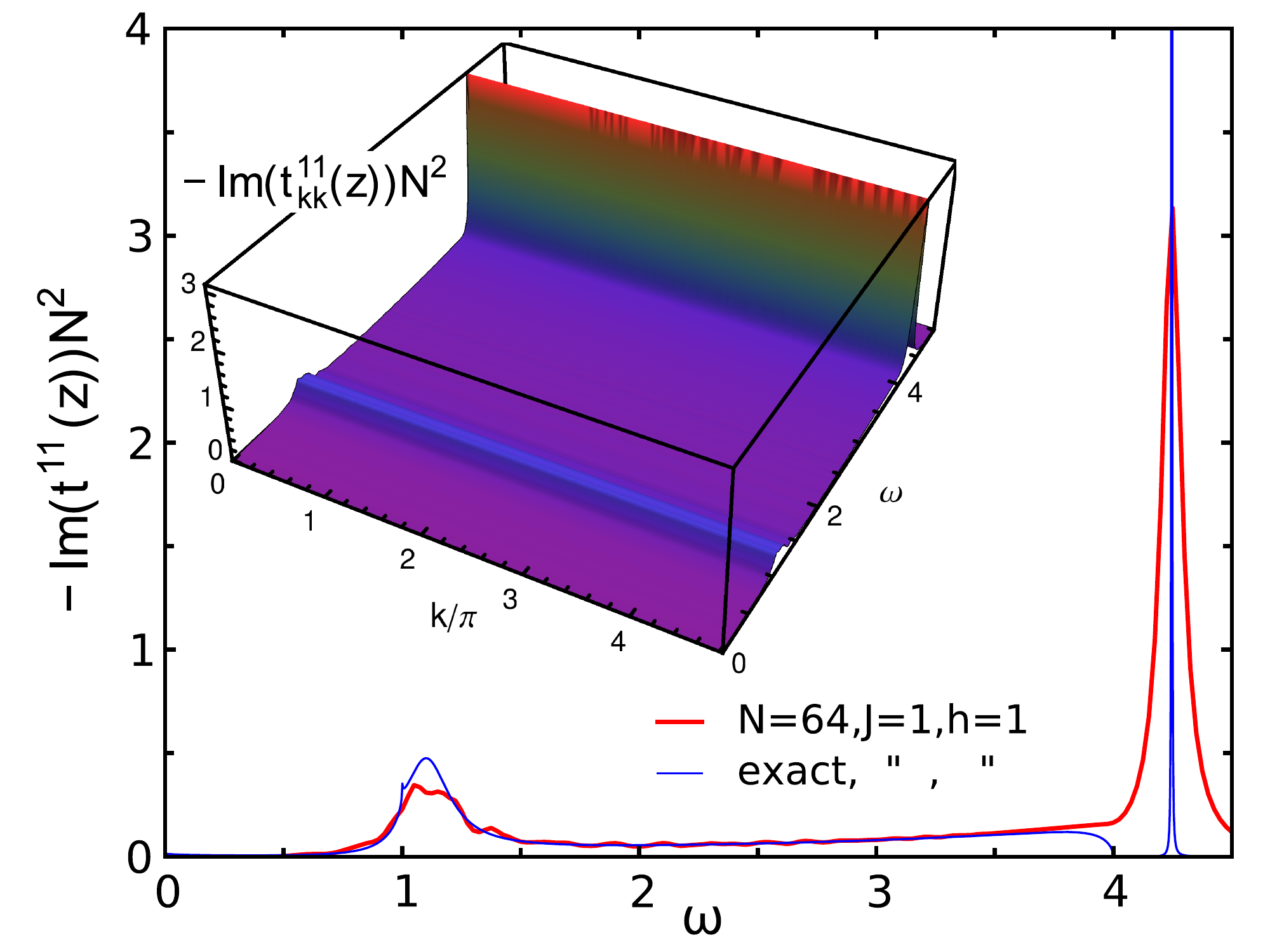}
\caption{\label{fig3}(color online) Analytical and numerical results for the $T$-matrix spectrum in
no-texture case, $J\!=\!1$, $h\!=\!1$. Homogeneous canting angles of the host spins $\theta\!\simeq\! 0.1253$.
Impurity canting angle $\theta_i\! \simeq\! 0.8729$ as in the actual texture.  Thin
solid blue: exact $-{\rm Im}\,t^{11}(z{=}\omega)$ \cite{supp}. Impurity resonance energy at
$\varepsilon\!\simeq\!1.308$, anti-bound state at $\simeq\!4.246$, van Hove singularities at $\omega_{gap}\!=\! 1$
and  $\omega^{max}\!\approx\!4$ are clear. Thick red solid: numerical
$-{\rm Im}\,t^{11}_{{\bf k}'{=}{\bf k}{=}0}(z{=}\omega{+}i 0.05)$ for $N{=}64$. Inset: numerical
$-{\rm Im}\,t^{11}_{{\bf k},{\bf k}}(z{=}\omega{+}i 0.05)$ along the ${\bf k}$-path of Fig. \ref{fig4}. }
\end{figure}

\emph{No texture test.}---First, we demonstrate the feasibility of obtaining
the $T$-matrix from Eq. (\ref{eq6}) numerically.  For that purpose, we solve a
complementary \emph{artificial} problem, in which we neglect the feedback of the
impurity on the host spins, i.e., spins in the plane retain their homogeneous
field-induced canting of Fig.~\ref{fig1}(b) and no texture is created.  While such a
reference state is, of course, unstable as ${\cal H}_{\mathrm{class}}$ is not at its
minimum, it permits an analytical solution of the scattering problem of ${\cal H}_2$,
details of which are provided in \cite{supp}. The analytical solution can be compared to
the $T$-matrix obtained from the numerical procedure
described above. In the following we consider the resolvent, i.e., the $T$-matrix
stripped from the matrices of the Bogolyubov basis transformation
\begin{equation}
\mathbf{t}_{\mathbf{k'k}}(z)=(\mathbf{B}_{\mathbf{k'}}^{\dagger})^{-1}
\mathbf{T}_{\mathbf{k'k}}(z)(\mathbf{B}_{\mathbf{k}})^{-1}\,,
\label{resolve}
\end{equation}
where $B^{11(22)}_{\bf k}{=}u_{\bf k}$ and $B^{12(21)}_{\bf k}{=}v_{\bf k}$
\cite{supp}.

\begin{figure}
\includegraphics[width=0.55\columnwidth]
{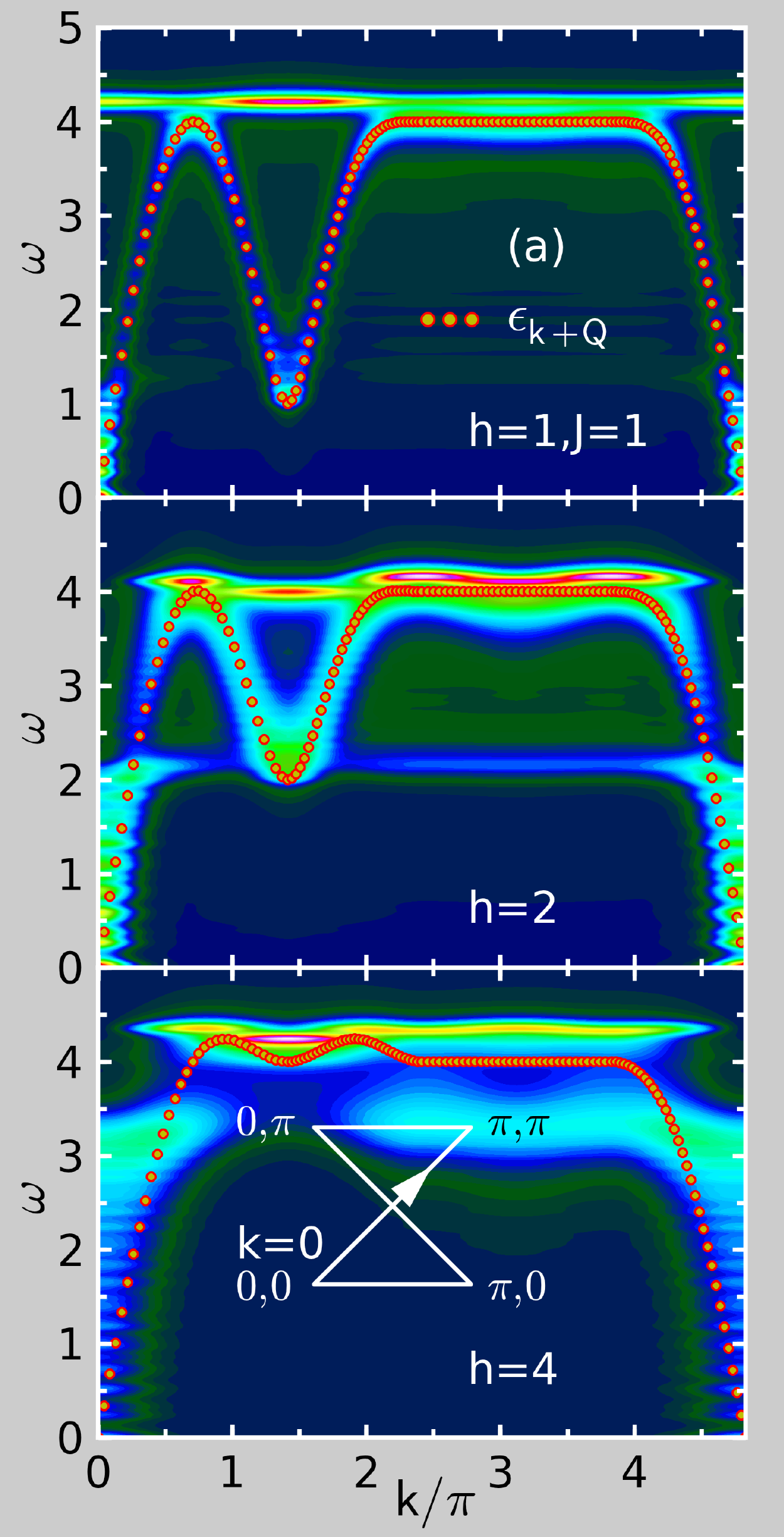}~~~\includegraphics[width=0.42\columnwidth]{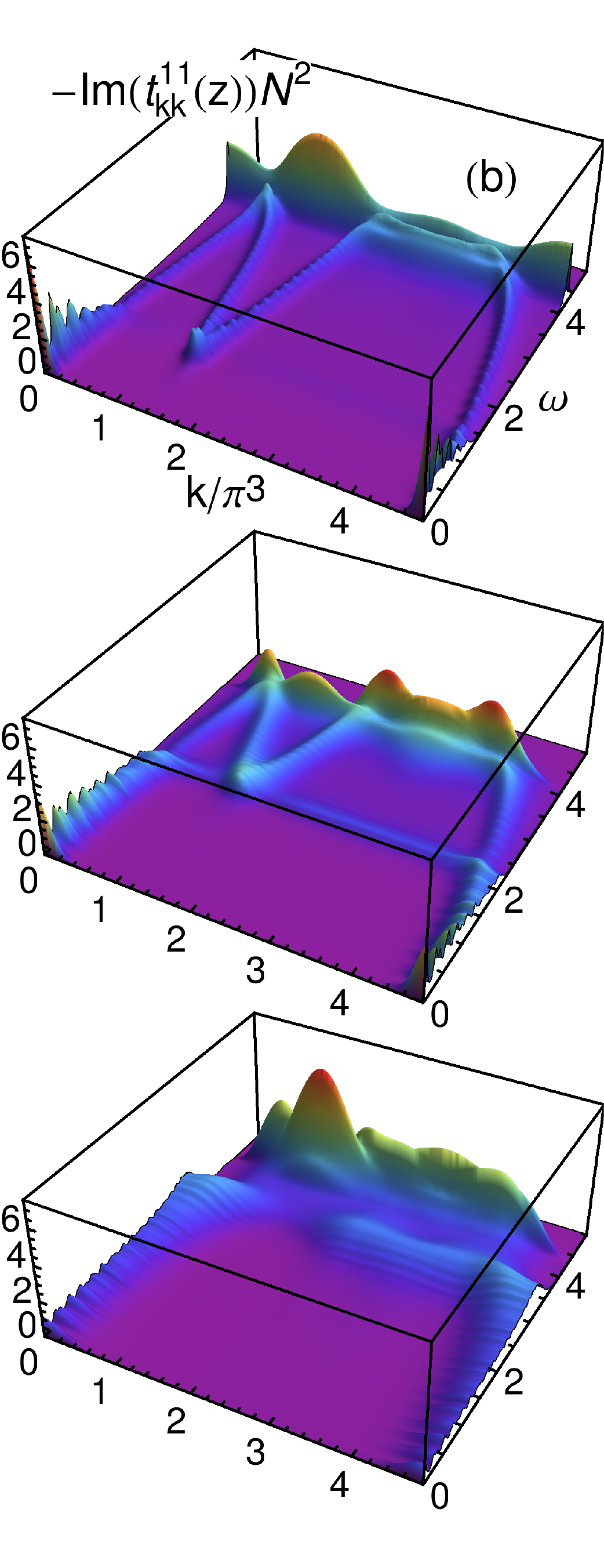}
\caption{\label{fig4} (color online) The $T$-matrix resolvent spectra
$-{\rm Im}\,t^{11}_{{\bf k},{\bf k}}(z{=}\omega{+}i 0.05)$ in $N\!=\!64$ cluster vs ${\bf k}$ and $\omega$,
$J\!=\!1$, $h{=}1$, 2, and 4
along the depicted ${\bf k}$-path. Panel (a): contour plots superimposed with the shifted magnon
dispersion $\varepsilon_{{\bf k}+{\bf Q}}$ (red-yellow dots). Panel (b): 3D plots.}
\end{figure}

The analytical result for the the resolvent spectrum, $-\textrm{Im}\,
t^{11}_{{\bf k}'{\bf k}}(z)$, is plotted in Fig. \ref{fig3} vs frequency
$\omega$.  Naturally, $\mathbf{t}_{\mathbf{k'k}}(z) \equiv \mathbf{t}(z)$ is
{\em momentum independent} \cite{supp}. This is an expected behavior for
scattering from point-like defects and is similar to scattering from vacancies
in collinear HAFs \cite{Brenig91,Chernyshev01}, where the resolvent shows some
broad ${\bf k}$-modulation from superposition of a small number of partial
waves. The inset of Fig.~\ref{fig3} shows
$-\textrm{Im}\,t^{11}_{\mathbf{kk}}(z)$ obtained numerically from (\ref{eq6})
and (\ref{resolve}) along the path in ${\bf k}$-space shown in
Fig.~\ref{fig4}. Clearly, it is also momentum independent. In addition,
analytical and numerical results, if evaluated on the finite clusters of the
same size, agree to within numerical precision \cite{supp}.

Finally, Fig.~\ref{fig3} demonstrates the spectral resolution we can obtain from
the numerical procedure in an $N\!=\!64$ cluster with a minimally acceptable
imaginary broadening.  One can see, that the numerical scattering amplitude has
all the features of the analytical one: the impurity resonance, the shallow
spin-wave continuum, and the anti-bound state above the upper edge of the
spectrum \cite{Fulde95}.  Fine details, such as the anti-bound state gap and the
non-analytic van Hove singularities are smeared out.  Improving this with
systems sizes beyond $N \!=\! 70$ is impractical because of the large memory
requirements for the non-sparse $2(N^{2}\!+\!1){\times}2(N^{2}\!+\!1)$ matrices.

\emph{Dispersive resonance.}---We now consider the scattering $T$-matrix for the true
ground state of the system with the spin texture. An analytical solution is not possible
in this case.  With the feasibility of the numerical procedure established, we evaluate
the $T$-matrix from Eq. (\ref{eq6}) using $\theta_{\{l\}}$ from the minimization of
${\cal H}_{\mathrm{class}}$ as an input to the Bogolyubov transformation. Representative
results are shown in Fig.~\ref{fig4}.  Removing the ${\bf k}$-dependence due to
transformation of the basis from $\mathbf{T}_{\mathbf{kk}}(z)$ as in (\ref{resolve}), we
show $-\textrm{Im}\, t^{11}_{{\bf k}{\bf k}}(z)$ as a function of $\omega$ and
${\bf k}$ along a high-symmetry path in the Brillouin zone and for several values of the
magnetic field.

In a sharp contrast to the no-texture case, $\textrm{Im}\,t_{\mathbf{kk}}^{11}(z)$
reveals a clear dispersive feature. The localized impurity resonance  in Fig.~\ref{fig3}
is now visible only as a faint  maximum and is completely
overshadowed by the dispersive resonance.  Such a result is completely unexpected for
the point-like impurity coupled to the Heisenberg model (\ref{eq1}). Direct comparison
in Fig.~\ref{fig4}(a) shows that the ${\bf k}$-dependence of the dispersive resonance
closely follows the spinwave dispersion $\varepsilon_{\mathbf{k}+\mathbf{Q}}$, folded by
the ordering vector $\mathbf{Q}=(\pi,\pi)$. As one can see, the resonance is most
sharply defined for small fields and gets washed out at higher fields. We find the
dispersive feature to be prominent regardless of the system size or the impurity
coupling $J$.

It is reasonable to suggest that the dispersive resonance is a natural outcome of the
scattering from an extended region of the impurity-induced texture, arising due to
non-collinearity of the state. This can be understood qualitatively from
Fig.~\ref{fig1}(c), which shows that the impurity spin has a component that acts as a
local field in the direction perpendicular to the homogeneous field-induced
canting. Because of that, the spins of Fig.~\ref{fig1}(b)
are perturbed from  their local reference frames by the \emph{staggered} transverse effective field.  Then the spin-wave
part of the Hamiltonian can be written as ${\cal H}_{2}={\cal H}^{h,i}+{\cal H}^{stag}$,
where ${\cal H}^{h,i}$ contains the homogeneous canting of spins and the point-like
impurity scattering as in the no-texture case, while ${\cal H}^{stag}$ is inhomogeneous
with \emph{staggered} matrix elements, which decay on the length scale set by the
texture.

Because of the staggering, magnons must experience an umklapp scattering potential that
can be approximated, for an extended region of the texture, as ${\cal
  H}^{stag}\sim\sum_{\bf k} \mathbf{W}_{\bf k} b_{\bf k+Q}^\dag b_{\bf
  k}^{\phantom{\dagger}}$.  Here, a qualitative analogy can be drawn with the 1D
Kronig-Penney model whose $T$-matrix is dispersive and has a pole close to the
zone-folded band $\varepsilon_{k+Q}$ \cite{Marder}. Because of the finite spatial extent of
the texture, the dispersive resonance must be broadened. This is consistent with the
increase of the broadening in Fig.~\ref{fig4} at higher fields where the size of the
texture shrinks.  This may imply a nontrivial behavior of the $T$-matrix in the limit of
$h\rightarrow 0$ where the texture becomes quasi-long-ranged as in  magnets that are
non-collinear in zero field. We note that the impurity scattering does not lead to
overdamping of the Goldstone mode, i.e., the spectral density at low energies  in
Fig.~\ref{fig4} does not occur at the ordering vector $\mathbf{Q}$.

Our results are of a direct relevance to the excitation spectra of non-collinear magnets
with a low concentration $x$ of naturally occurring or deliberately doped impurities.
Since the magnon self-energy is simply proportional to the diagonal element of the
$T$-matrix via $\mathbf{\Sigma}_{\bf k}(\omega)\sim x \mathbf{T}_{\mathbf{kk}}(\omega)$,
one may expect to observe an anomalous ${\bf k}$-dependent broadening of the spectrum
where $\varepsilon_{\mathbf{k}}$ overlap with $\varepsilon_{\mathbf{k}+\mathbf{Q}}$ and an
equally unusual field-dependence of such a broadening. These and other features should
be observable by inelastic neutron scattering and specific predictions will be subject
of future work.

{\it Conclusions.}---%
To conclude, we have presented strong evidence for a highly anomalous static and
dynamic response of non-collinear antiferromagnets to doping by point-like defects.
The scattering amplitude exhibits features that are strikingly different from the usual
$s$-wave scattering and include a highly dispersive resonance due to an impurity-induced
texture.  This result should be valid for the broad class of non-collinear magnets.
Further theoretical and experimental studies seem highly desirable.

Part of this work has been done at the Kavli Institute for Theoretical Physics  (A.L.C. and W.B.) and at the
Platform for Superconductivity and Magnetism, Dresden (W.B.).
The work of A.L.C. was supported by the DOE under Grant No. DE-FG02-04ER46174. The work of W.B.
was supported by DFG FOR912 Grant No. BR 1084/6-2, EU MC-ITN LOTHERM Grant No. PITN-GA-2009-238475,
and the NTH SCNS. The research at KITP was supported by NSF Grant No. NSF PHY11-25915.

\onecolumngrid
\newpage
\newpage
\setcounter{page}{1}
\begin{center}
{\large\bf Highly Dispersive Scattering From Defects In Non-Collinear Magnets:\\
Supplemental Information}\\
\vskip0.35cm
Wolfram Brenig$^{1}$ and A. L. Chernyshev$^{2}$\\
\vskip0.15cm
{\it \small
$^1$Institut f\"{u}r Theoretische Physik, Technische Universit\"{a}t Braunschweig,
38106 Braunschweig, Germany\\
$^2$Department of Physics, University of California, Irvine, California 92697, USA}\\
{\small (Dated: November 19, 2012)}\\
\end{center}
\twocolumngrid

\begin{figure}[b]
\includegraphics[width=0.6\columnwidth]{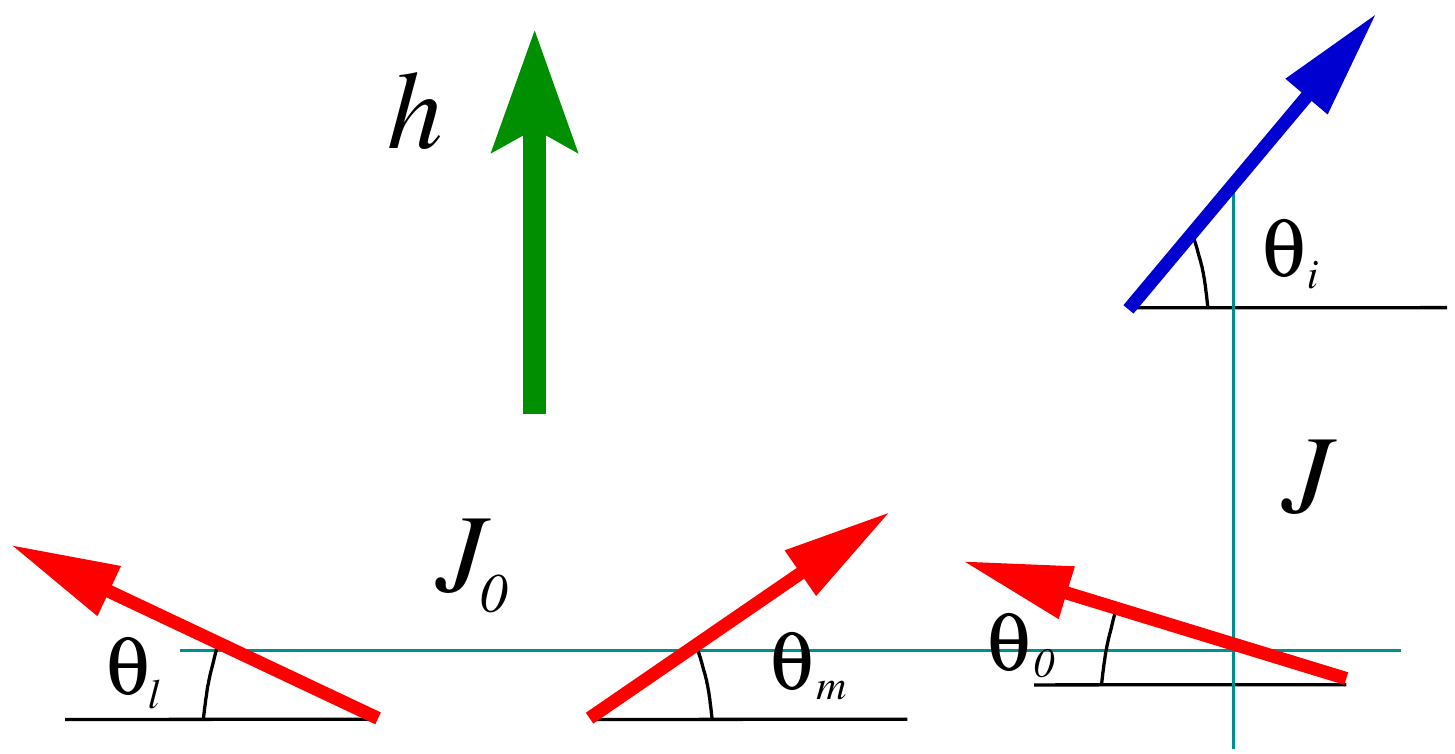}
\caption{\label{fig5}(color online) A sketch of the spin configuration.}
\end{figure}

\subsection{Classical Hamiltonian}

\begin{figure}[t]
\noindent \centering{}\includegraphics[width=0.8\columnwidth]{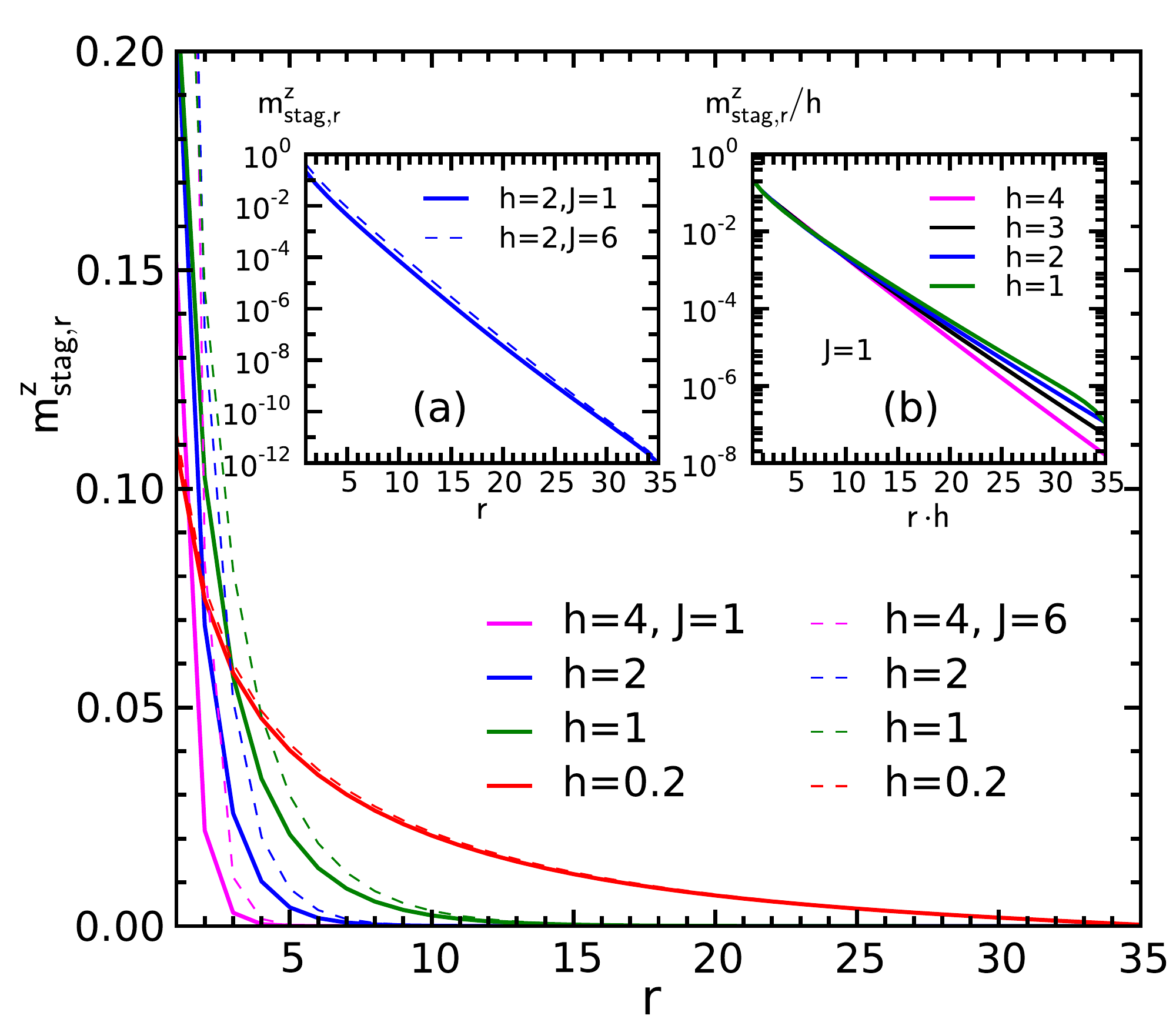}
\caption{\label{fig6}(color online) The staggered component of magnetization
$m_{stag,{\bf r}}^{z}$ (\ref{mstagg}) vs distance
$r$ from the origin for two different values of impurity coupling $J$ and for
various field strengths.  The impurity is coupled to site $R=(0,0)$ of a
cluster with $N=70$.  Inset (a) same on the semi-log plot. Inset (b)
$m_{stag,{\bf r}}^{z}/h$ vs $r\cdot h$.}
\vskip -0.2cm
\end{figure}

The classical energy of the square-lattice Heisenberg AF in a field with an
out-of-plane impurity, model (1) of the main text, is
\begin{eqnarray}
{\cal H}_{\mathrm{class}}&=&-S^2J_0\sum_{\langle lm\rangle}\cos(\theta_{l}+\theta_{m})
-Sh\sum_{l} \sin\theta_{l}
\nonumber \\
&&-JSS^\prime\cos(\theta_{0}+\theta_{i}) -hS^{\prime}\sin\theta_i\, ,
\label{Hclass}
\end{eqnarray}
where $\langle lm\rangle$ denotes bonds. The impurity site $i$ is coupled to
site $l=0$ of the host and $\theta_n$'s are the tilt angles out of the
$xy$-plane, see Fig.~\ref{fig5}.  The classical ground state is obtained by
numerical minimization of ${\cal H}_{\mathrm{class}}$, i.e. by
solving the set of equations $\partial {\cal
H}_{\mathrm{class}}/\partial\theta_{\{l\}}=0$.  The resulting spin configuration
corresponds to an \emph{inhomogeneous} distribution of spin canting,
parametrized by the local tilt angles $\theta_l$, i.e. the \emph{texture}.  Our
Fig.~\ref{fig6} exhibits one of the quantities that can be used to analyze the
spatial extent and other characteristics of the texture: the staggered component
of the magnetization $m_{stag,{\bf r}}^{z}$.
The main panel displays
$m_{stag,{\bf r}}^{z}$ along
the $x$-axis, at location ${\bf r}=(r,0)$ off the impurity, where
\begin{eqnarray}
m_{stag,{\bf r}}^{z}=(-1)^r\,\frac{\left(S_{(r+1,0)}^{z}-S_{(r,0)}^{z}\right)}{2}\,,
\label{mstagg}
\end{eqnarray}
for two different values of impurity coupling $J$ and for various field
strengths.  The impurity is coupled to site $R=(0,0)$ of a cluster with $N=70$.
Clearly, the spatial extent of the texture increases as the field
$h\!\rightarrow\!0$.  Inset (a) demonstrates that the texture decays
exponentially \cite{commentAccuracy} at $r\gg1$. This behavior is expected
because the impurity is not coupled to the Goldstone mode of the host. This
finding is consistent with earlier work \cite{Eggert07a} where $m_{stag,{\bf
r}}^{z}$ was investigated for a different type of impurity (vacancy) by a
continuum theory and quantum Monte Carlo. We do not observe exact exponential
behavior in the low-field regime, most likely because of the finite cluster
size. By varying $h$ we find that $m_{stag,{\bf r}}^{z}/h$ scales almost
perfectly with $r\cdot h$, see inset (b) of Fig.~\ref{fig6}, again in agreement
with \cite{Eggert07a}. One potential reason for the visible deviation from scaling
is the field dependence of the transverse susceptibility $\chi_{\perp}$
and the spin stiffness $\rho_{s}$, neglected in the continuum description of
\cite{Eggert07a}.

\subsection{Harmonic part of the Hamiltonian}

Within the $1/S$ expansion, the local spin quantization axes on each site are
aligned in the direction given by the local canted frame with the angle
$\theta_l$, obtained from the minimization of the classical energy. The
subsequent Holstein-Primakoff bosonization of spin operators yields the harmonic
spin-wave Hamiltonian ${\cal H}_{2}$.  The term linear in bosonic operators,
${\cal H}_{1}$, vanishes identically upon minimization of ${\cal
H}_{\mathrm{class}}$.

The spin-wave Hamiltonian of the host reads
\begin{eqnarray}
&&{\cal H}^{\rm host}_{2}=J_0S\sum_{l,\delta} \bigg\{\cos\theta_{l\delta}\ a^\dag_l a^{\phantom{\dag}}_l
+\sin^2\frac{\theta_{l\delta}}{2}\ a^\dag_l a^{\phantom{\dag}}_{l+\delta}
\label{eq2}
\\
&&\phantom{{\cal H}_2}
-\frac12 \, \cos^2\frac{\theta_{l\delta}}{2}\left( a^\dag_l a^{\dag}_{l+\delta}+{\rm H.c.}
\right)\bigg\}
+h\sum_l\sin\theta_{l}\ a^\dag_l a^{\phantom{\dag}}_l\, ,
\nonumber
\end{eqnarray}
where the summation is over the lattice sites $l$ and the nearest neighbors
$\delta$ and the shorthand notation
$\theta_{l\delta}=\theta_{l}+\theta_{l+\delta}$ has been introduced.
The impurity part
of the Hamiltonian is
\begin{eqnarray}
{\cal H}^{\rm imp}_{2}&=&J\cos\theta_{0i}\left(S\, a^\dag_0 a^{\phantom{\dag}}_0+
S'\, a^\dag_i a^{\phantom{\dag}}_i\right)+h\,\sin\theta_i\ a^\dag_i a^{\phantom{\dag}}_i\nonumber\\
&+&
J\sqrt{SS'}\bigg\{\sin^2\frac{\theta_{0i}}{2}\left( a^\dag_0 a^{\phantom{\dag}}_i+{\rm H.c.}\right)
\label{eq2a}
\\
&&\phantom{J\sqrt{SS'}\bigg\{}-
 \cos^2\frac{\theta_{0i}}{2}\left( a^\dag_0 a^{\dag}_i+{\rm H.c.}
\right)\bigg\} ,
\nonumber
\end{eqnarray}
where $\theta_{0i}=\theta_0+\theta_i$. The first line contains a potential-like
energy shift for the magnon on the site coupled to the impurity ($l=0$) and the
local energy of the impurity magnon, while the rest of the Hamiltonian describes
various transitions between the two.

In the textured case, analytical diagonalization of the Hamiltonian (\ref{eq2}) and
(\ref{eq2a}) is not feasible, so we perform the Bogolyubov transformation
numerically. Subsequently, the $T$-matrix is extracted from the Green's function
written in the basis of Bogolyubov magnons of the {\it uniform} system (no
impurity), as described in the main text.

\subsection{No texture test}

The feasibility of extracting the $T$-matrix from the full Green's function
numerically can be demonstrated for a complementary \emph{artificial} problem,
for which an analytical solution can also be found, and by comparing the results of
the two approaches. Here we formulate such a problem by neglecting the feedback
of the impurity onto the host spins, so that no texture is created.  While such a
reference state is unstable as ${\cal H}_{\mathrm{class}}$ is not minimal, it
permits an analytical solution of the scattering problem.

\subsubsection{Hamiltonian}

In the no-texture case, the canting of the host spins is uniform,
$\theta_l=\theta$, with the canting angle found from the energy minimization of
the system without impurity: $\sin\theta=h/h_{s}$ with the saturation field
$h_{s}=8SJ_0$.  The subsequent diagonalization of the host Hamiltonian
(\ref{eq2}) is straightforward \cite{Zhitomirsky98a,Mourigal10a} and leads to
\begin{eqnarray}
{\cal H}^{\rm
host}_{2}=\sum_{\bf{k}}\varepsilon_{\bf{k}}b_{\bf{k}}^{\dagger}
b_{\bf{k}}^{\phantom{\dagger}},
\label{eq3a}
\end{eqnarray}
where $\varepsilon_{\bf{k}}=4SJ_0\sqrt{(1+\gamma_{\bf{k}})(1-\gamma_{\bf{k}}\cos2\theta)}$
is the spin-wave dispersion
and $\gamma_{\bf{k}}=\frac12(\cos k_{x}+\cos k_{y})$. The  Bogolyubov transformation
from $a_{\bf k}^{\dagger} =\frac{1}{N}\sum_{l}
e^{i{\bf k}\cdot\mathbf{r}_{l}}a_{l}^{\dag}$
to $b^{\phantom{\dag}}_{\bf{k}}$ and $b_{\bf{k}}^{\dagger}$ is written as
\begin{eqnarray}
  \Phi_{\bf k}=\mathbf{B_{\bf k}}\Psi_{\bf{k}}\, ,
\label{eq3b}
\end{eqnarray}
where
\begin{eqnarray}
\label{psi_n_B}
\Phi_{\bf k}=\left[\begin{array}{c}
a_{\bf k}^{\phantom{\dag}}\\ a_{-\bf k}^{\dag}
\end{array}\right] ,  \ \
\Psi_{\bf{k}}=\left[\begin{array}{c}
b_{\bf k}^{\phantom{\dag}}\\ b_{-\bf k}^{\dag}
\end{array}\right] , \ \
 \mathbf{B_{\bf k}}=\left[\begin{array}{cc}
u_{\bf k} & v_{\bf k}\\
v_{\bf k} & u_{\bf k}\end{array}\right], \ \ \ \ \ \
\end{eqnarray}
is a convenient $2\times 2$ notation.  The $u$-$v$ factors are: $u_{\bf
k}=\sqrt{(A_{\bf k}+\varepsilon_{\bf k})/2\varepsilon_{\bf k}}$, $v_{\bf
k}=\mathrm{sign}(\gamma_{\bf k})\sqrt{(A_{\bf k}-\varepsilon_{\bf
k})/2\varepsilon_{\bf k}}$, with $A_{\bf k}=4SJ_0(1+\gamma_{\bf
k}\sin^{2}\theta)$ \cite{Zhitomirsky98a,Mourigal10a}.

Using these notations and that ${\bf B}_{\bf k}={\bf B}_{\bf k}^\dag$, the
impurity part of the spin-wave Hamiltonian (\ref{eq2a}) can be written as
\begin{eqnarray}
{\cal H}^{\rm imp}_{2}&=&\varepsilon\, a_{i}^{\dag}a_{i}^{\phantom{\dag}}+
\frac{V_0}{2} \sum_{\bf k,k'}\Psi_{\bf k'}^{\dag}{\bf B}_{\bf k'}{\bf B}_{\bf k}\Psi_{\bf k}^{\phantom{\dag}}
+\sum_{\bf k}\Psi_{\bf k}^{\dagger}{\bf B}_{\bf k}{\bf V}_1\Phi_{i}^{\phantom{\dagger}},\nonumber\\
{\rm with} &&\Phi_{i}=\left[\begin{array}{c}
a_{i}^{\phantom{\dag}}\\ a_{i}^{\dag}
\end{array}\right] , \ \ \  \
{\bf V}_1=V_1\left[\begin{array}{cc}
f & g\\
g & f
\end{array}\right],
\label{eq3}
\end{eqnarray}
where potential-like scattering amplitude $V_0\!=\!JS\cos\theta_{0i}$,
magnon-impurity transfer amplitude $V_1\!=\!J\sqrt{SS'}$,  matrix
elements $f\!=\!\sin^2(\theta_{0i}/2)$ and $g\!=\!-\cos^2(\theta_{0i}/2)$, and  the
angle $\theta_{0i}\!=\!\theta\!+\!\theta_{i}$ as before. With the notations
(\ref{psi_n_B}), the last term of the magnon-impurity scattering in (\ref{eq3})
implicitly contains its own conjugate. The impurity energy is
$\varepsilon\!=\!JS'\cos\theta_{0i}\!+\!h\sin\theta_{i}$ with an impurity
canting angle $\theta_i$ which is a free parameter. In the following, we fix the
latter to its value obtained numerically for the problem with the texture. One
might also chose $\theta_i$ to minimize the energy of the impurity spin in
(\ref{Hclass}) while keeping the host canting angle homogeneous. This yields
$\tan\theta_i\!=\!(8J_0/J\!-\!1)\tan\theta$, which is numerically close to the
problem with the texture.

\subsubsection{Green's function and $T$-matrix}

\begin{figure}[t]
\includegraphics[width=0.8\columnwidth]{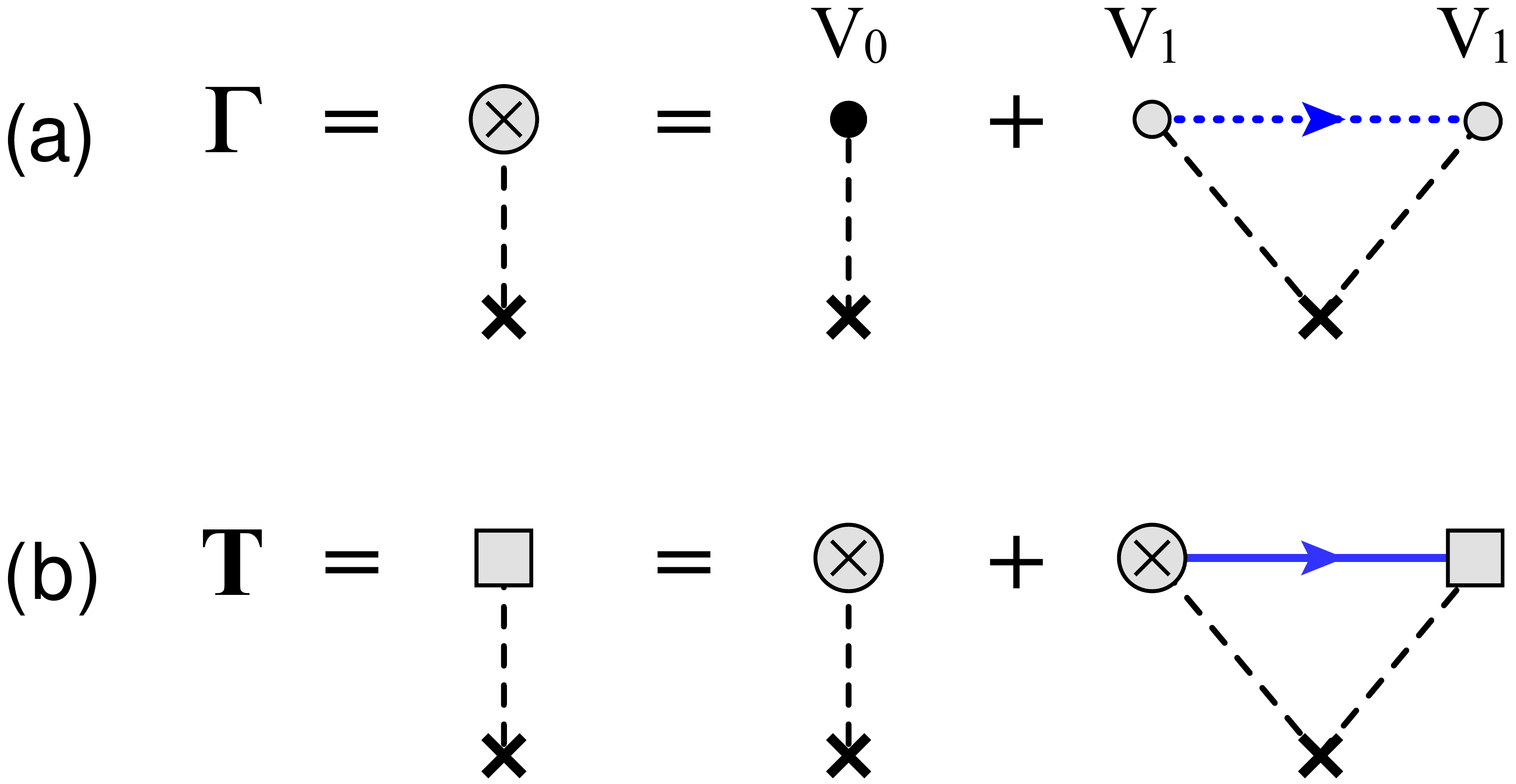}
\caption{\label{fig7}(color online) (a) Composite vertex of impurity scattering in
(\ref{eq3}). (b) $T$-matrix sequence with that vertex. Dotted and solid lines
are the noninteracting impurity resonance and the host magnon Green's functions,
respectively.}
\end{figure}

The imaginary time Green's function of the $b^{(\dag)}_{\bf{k}}$
magnons is a $2\!\times\!2$ matrix which can be written as
a direct product ${\bf G}_{\bf k'\bf k}(\tau)\!=\!-\langle T_{\tau}(\Psi_{\bf
k'}^{\phantom{\dagger}}(\tau)\!\otimes\!\Psi_{\bf k}^{\dagger})\rangle$ using
(\ref{psi_n_B}). Its Fourier transform is ${\bf G}_{\bf k'\bf
k}(i\omega_{n})\!=\!\int_{0}^{\beta}\exp(i\omega_{n}\tau){\bf G}_{\bf k'\bf
k}(\tau)d\tau$, where $i\omega_{n}\!=\!i2n\pi T$, which we replace by a complex
variable $z$ hereafter. The Green's function in the presence of impurity
scattering can be expressed through the $T$-matrix
\begin{eqnarray}
{\bf G}_{\bf k'\bf k}(z) & = & \delta_{\bf k'\bf k}{\bf G}_{\bf k}^{0}(z)+
{\bf G}_{\bf k'}^{0}(z){\bf T}_{\bf k'\bf k}(z){\bf G}_{\bf k}^{0}(z)\, ,
\label{eq4}
\end{eqnarray}
where the noninteracting Green's function ${\bf G}_{\bf k}^{0}(z)$ is
\begin{eqnarray}
{\bf G}_{\bf k}^{0}(z) & = & \left[\begin{array}{cc}
 \frac{1}{z-\varepsilon_{\bf k}} & 0\\
 0 & -\frac{1}{z+\varepsilon_{\bf k}}\end{array}\right]\, .
\end{eqnarray}
Using the impurity scattering terms in (\ref{eq3}), the $T$-matrix follows from
an infinite sequence of the two-component vertex function, one component from
the potential-like scattering $V_0$-term and the other from the magnon-impurity
scattering $V_1$-term.  Figs.~\ref{fig7}(a) and (b) show the vertex and the
sequence, respectively.  From the structure of the vertex
\begin{eqnarray}
\mathbf{\Gamma}_{\bf k'\bf k}(z)  =  {\bf B}_{\bf k'}^{\dagger}\mathbf{\Gamma}(z)
{\bf B}_{\bf k}\, ,
\label{Gamma}
\end{eqnarray}
where the ${\bf k}$-independent vertex is
\begin{eqnarray}
\mathbf{\Gamma}(z)  = V_0{\bf 1}+{\bf V}_1{\bf G}_{i}^{0}(z){\bf V}_1\, ,
\label{Gamma1}
\end{eqnarray}
and ${\bf G}_{i}^{0}(z)$ is the local Green's function of the impurity state,
\begin{eqnarray}
{\bf G}_{i}^{0}(z)=\left[\begin{array}{cc}
\frac{1}{z-\varepsilon} & 0\\
0 & -\frac{1}{z+\varepsilon}
\end{array}\right],
\label{Gamma2}
\end{eqnarray}
it is straightforward to see that the $T$-matrix has no ${\bf k}$-dependence
aside from the trivial basis change ${\bf B}_{\bf k}$ and ${\bf B}_{\bf
k^\prime}$ to Bogolyubov bosons
\begin{eqnarray}
{\bf T}_{\bf k'\bf k}(z)  =  {\bf B}_{\bf k'}^{\dagger}{\bf t}(z){\bf B}_{\bf k}\, .
\label{T1}
\end{eqnarray}
This is a standard feature, not only for the particular type of impurity in
antiferromagnets that we investigate, but for the  point-like scatterers in general.
Stated differently, the actual scattering resolvent
${\bf t}(z)$ is completely \emph{momentum independent}.  Using (\ref{Gamma}),
(\ref{Gamma1}), and Fig.~\ref{fig7}(b) one readily arrives at the final expression
for the resolvent
\begin{eqnarray}
{\bf t}(z)=\frac{\mathbf{\Gamma}(z)}{{\bf 1}-{\bf F}(z)\mathbf{\Gamma}(z)}\, ,
\label{t1}
\end{eqnarray}
where the matrix ${\bf F}(z)$
\begin{eqnarray}
{\bf F}(z) =
\left[\begin{array}{cc}
G_{0}(z) & F_{0}(z)\\
F_{0}(z) & G_{0}(-z)
\end{array}\right],
\label{eq:5}
\end{eqnarray}
is built from the local Holstein-Primakoff Green's functions of the host
\begin{eqnarray}
G_{0}(z) & = & \sum_{\bf k}\left(\frac{u_{\bf k}^{2}}{z-\varepsilon_{\bf k}}-
\frac{v_{\bf k}^{2}}{z+\varepsilon_{\bf k}}\right)\nonumber \\
F_{0}(z) & = & \sum_{\bf k}\left(\frac{v_{\bf k}u_{\bf k}}{z-\varepsilon_{\bf k}}-
\frac{u_{\bf k}v_{\bf k}}{z+\varepsilon_{\bf k}}\right)\,.
\label{eq:5a}
\end{eqnarray}
This concludes the formal solution of the scattering problem. The remaining step is the momentum
integration in $G_0(z)$ and $F_0(z)$. In the next section, we summarize our
{\em analytical} results for this integration.

\subsubsection{Local Green's functions}

Since in the scattering problem we are interested in the retarded $T$-matrix, we need to evaluate
$G_0(z)$ and $F_0(z)$ in (\ref{eq:5a}) for $z=\omega+i0^+$. Both Green's functions
have real and imaginary parts:
\begin{eqnarray}
G_{0}(\omega) & = & G'_{0}(\omega)+iG''_{0}(\omega)\nonumber \\
F_{0}(\omega) & = & F'_{0}(\omega)+iF''_{0}(\omega)\,.
\label{G0s}
\end{eqnarray}
It is convenient to express all energies in units of the zero-field magnon bandwidth, $W_0=4J_0S$,  and the field in units of
the saturation field of the host, $h_s=8J_0S$. Thus, in the following $\bar{h}=h/h_s$ and $\bar{\omega}=\omega/4J_0S$.
Three other energy scales are needed for the results below: the field-dependent  gap in the magnon spectrum
\begin{eqnarray}
\bar{\omega}_{g}=\frac{\omega_{gap}}{4J_0S}=2\bar{h},
\label{wg}
\end{eqnarray}
an auxiliary scale
\begin{eqnarray}
\bar{\omega}_{m}=\frac{\omega_{m}}{4J_0S}=\frac{1-\bar{h}^2}{\sqrt{1-2\bar{h}^2}},
\label{wm}
\end{eqnarray}
and the field-dependent magnon bandwidth
\begin{eqnarray}
\frac{W}{4J_0S}=\begin{cases}
\bar{\omega}_{m}, & \bar{h}\leq 1/\sqrt{3} \\
\bar{\omega}_{g}, & \bar{h}\geq 1/\sqrt{3} \, .
\end{cases}
\label{W}
\end{eqnarray}

For $-W<\omega<W$, and after some algebra, we arrive at the following expressions
for the imaginary parts
\begin{eqnarray}
&&G''_{0}(\omega)=\frac{-1}{\pi W_0r(\bar{\omega})}\bigg[\Theta\left(\bar{\omega}^2-\bar{\omega}_{g}^2\right)
 \left(1+\bar{\omega}+\bar{h}^2\gamma_1\right) K\left(\gamma'_1\right)
\nonumber \\
&&\phantom{G''_{0}(\omega)=\frac{-1}{\pi W_0r(\bar{\omega})}\bigg[}
+\left(1+\bar{\omega}+\bar{h}^2\gamma_2\right) K\left(\gamma'_2\right)\bigg],
\label{ImG0s}\\
&&F''_{0}(\omega)=\frac{-(1-\bar{h}^2)}{\pi W_0r(\bar{\omega})}
\bigg[\Theta\left(\bar{\omega}^2-\bar{\omega}_{g}^2\right)
\gamma_1 K\left(\gamma'_1\right)+ \gamma_2 K\left(\gamma'_2\right)\bigg]
.\nonumber
\end{eqnarray}
The expressions for the real parts of the local Green's functions for the same energy range $-W<\omega<W$ are
complimentary to (\ref{ImG0s})
\begin{eqnarray}
&&G'_{0}(\omega)=\frac{-1}{\pi W_0r(\bar{\omega})}\bigg[\Theta\left(\bar{\omega}^2-\bar{\omega}_{g}^2\right)
\left(1+\bar{\omega}+\bar{h}^2\gamma_1\right)K\left(\gamma_1\right)
\nonumber \\
&&\phantom{G'_{0}(\omega)=\frac{-1}{\pi}}
+\Theta\left(\bar{\omega}_{g}^2-\bar{\omega}^2\right)
\left(1+\bar{\omega}+\bar{h}^2\gamma_1\right)\frac{1}{\gamma_1} K\left(\frac{1}{\gamma_1}\right)
\nonumber\\
&&\phantom{G'_{0}(\omega)=\frac{-1}{\pi}}
-{\rm sign}\left(\gamma_2\right)\left(1+\bar{\omega}+\bar{h}^2\gamma_2\right) K\left(\gamma_2\right)\bigg],
\label{ReG0s} \\
&&F'_{0}(\omega)=\frac{-(1-\bar{h}^2)}{\pi W_0r(\bar{\omega})}\bigg[\Theta\left(\bar{\omega}^2-\bar{\omega}_{g}^2\right)
 \gamma_1K\left(\gamma_1\right)
\nonumber\\
&&\phantom{F'_{0}(\omega)=}
+\Theta\left(\bar{\omega}_{g}^2-\bar{\omega}^2\right) K\left(\frac{1}{\gamma_1}\right)
-{\rm sign}\left(\gamma_2\right) \gamma_2 K\left(\gamma_2\right)\bigg],\nonumber
\end{eqnarray}
where $\Theta$'s are step-functions, $K$'s are complete elliptic integrals of
the first kind, $\gamma'_i=\sqrt{1-\gamma_i^2}$, and
\begin{eqnarray}
\gamma_{1,2}=\frac{\bar{h}^2\pm r(\bar{\omega})}{1-2\bar{h}^2},
\label{gammas}
\end{eqnarray}
are the roots of the equation $\bar{\omega}^2-\bar{\omega}_{\bf k}^2=0$, i.e.
\begin{eqnarray}
r(\bar{\omega})=\sqrt{\left(1-2\bar{h}^2\right)\left(\bar{\omega}^2_m-\bar{\omega}^2\right)}\, .
\label{rw}
\end{eqnarray}
One may notice, that for the field $\bar{h}>1/\sqrt{3}$ some of the
step-functions [$\Theta\left(\bar{\omega}^2-\bar{\omega}_{g}^2\right)$] are zero
in the considered energy range $\omega^2<W^2$, because $W=\omega_{gap}$ for this
field range. As a consequence, some of the terms in (\ref{ImG0s}) and
(\ref{ReG0s}) vanish entirely for that field range.

For energies outside the magnon bandwidth, $\omega^2>W^2$, the imaginary parts
of $G_0$ and $F_0$ are identically zero. The derivation of the real parts, while
somewhat more convoluted, eventually leads to two cases.  First,
\begin{eqnarray}
&&G'_{0}(\omega)=-\frac{2}{\pi W_0r_1(\bar{\omega})}
{\rm Im}\bigg[\left(1+\bar{\omega}+\bar{h}^2\gamma_1\right)\frac{1}{\gamma_1} K\left(\frac{1}{\gamma_1}\right)\bigg],
\nonumber \\
&&F'_{0}(\omega)=-\frac{2(1-\bar{h}^2)}{\pi W_0r_1(\bar{\omega})}
{\rm Im}\bigg[ K\left(\frac{1}{\gamma_1}\right)\bigg],
\label{ReG0sa}
\end{eqnarray}
with
\begin{eqnarray}
r_1(\bar{\omega})=\sqrt{\left(1-2\bar{h}^2\right)\left(\bar{\omega}^2-\bar{\omega}_m^2\right)}\, .
\label{rw1}
\end{eqnarray}
This is valid for fields within the range $\bar{h}\leq 1/\sqrt{3}$ for any
$\omega^2>W^2$ and for $1/\sqrt{3}<\bar{h}\leq 1/\sqrt{2}$ for
$\omega^2>\omega^2_m$ ($\omega^2_m>W^2$).  Second,
\begin{eqnarray}
&&G'_{0}(\omega)=-\frac{1}{\pi W_0r(\bar{\omega})}\bigg[
\left(1+\bar{\omega}+\bar{h}^2\gamma_1\right)\frac{1}{\gamma_1} K\left(\frac{1}{\gamma_1}\right)
\nonumber \\
&&\phantom{G'_{0}(\omega)=-\frac{1}{\pi} W_0}
-\left(1+\bar{\omega}+\bar{h}^2\gamma_2\right)\frac{1}{\gamma_2} K\left(\frac{1}{\gamma_2}\right)\bigg],
\label{ReG0sb} \\
&&F'_{0}(\omega)=-\frac{(1-\bar{h}^2)}{\pi W_0r(\bar{\omega})}\bigg[K\left(\frac{1}{\gamma_1}\right)
-K\left(\frac{1}{\gamma_2}\right)\bigg],
\nonumber
\end{eqnarray}
which is valid for fields within the range $1/\sqrt{3}<\bar{h}\leq 1$ for any
$\omega^2>W^2$ and for the fields $1/\sqrt{3}<\bar{h}\leq 1/\sqrt{2}$ for
$\omega^2_m<\omega^2<W^2$.

\begin{figure}[b]
\includegraphics[width=1.0\columnwidth]{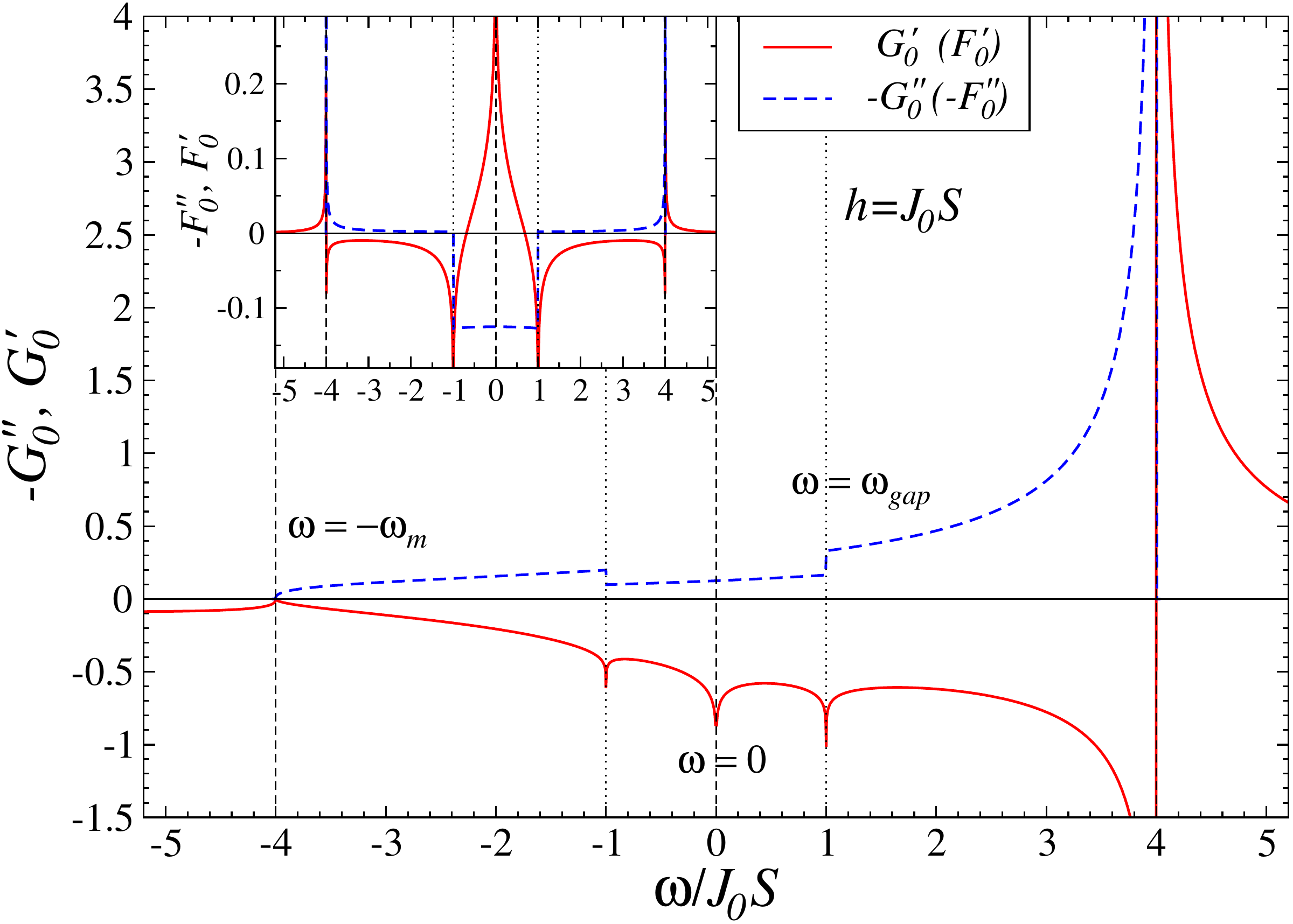}
\caption{\label{fig8}(color online) Real and imaginary parts of $G_0(\omega)$
and $F_0(\omega)$ for $h=J_0S$ ($\bar{h}=0.125$).}
\end{figure}

Fig.~\ref{fig8} shows the real and imaginary parts of $G_0$ and $F_0$ for a
representative choice of $h=J_0S$ ($\bar{h}=0.125$). Pronounced van Hove
singularities at the top of the magnon spectrum ($W\approx4J_0S$) and at the
field-induced gap ($\omega_{gap} = h$) are clearly visible. These nonanalyticities
are present in the $\omega$-dependence of the analytical result for
$-{\rm Im}\,t^{11}(\omega)$ from (\ref{t1}), which is depicted in Fig.~3 of the main
text, but are much less pronounced.

\subsubsection{Comparison of analytical and numerical results}

\begin{figure}[t]
\includegraphics[width=1.0\columnwidth]{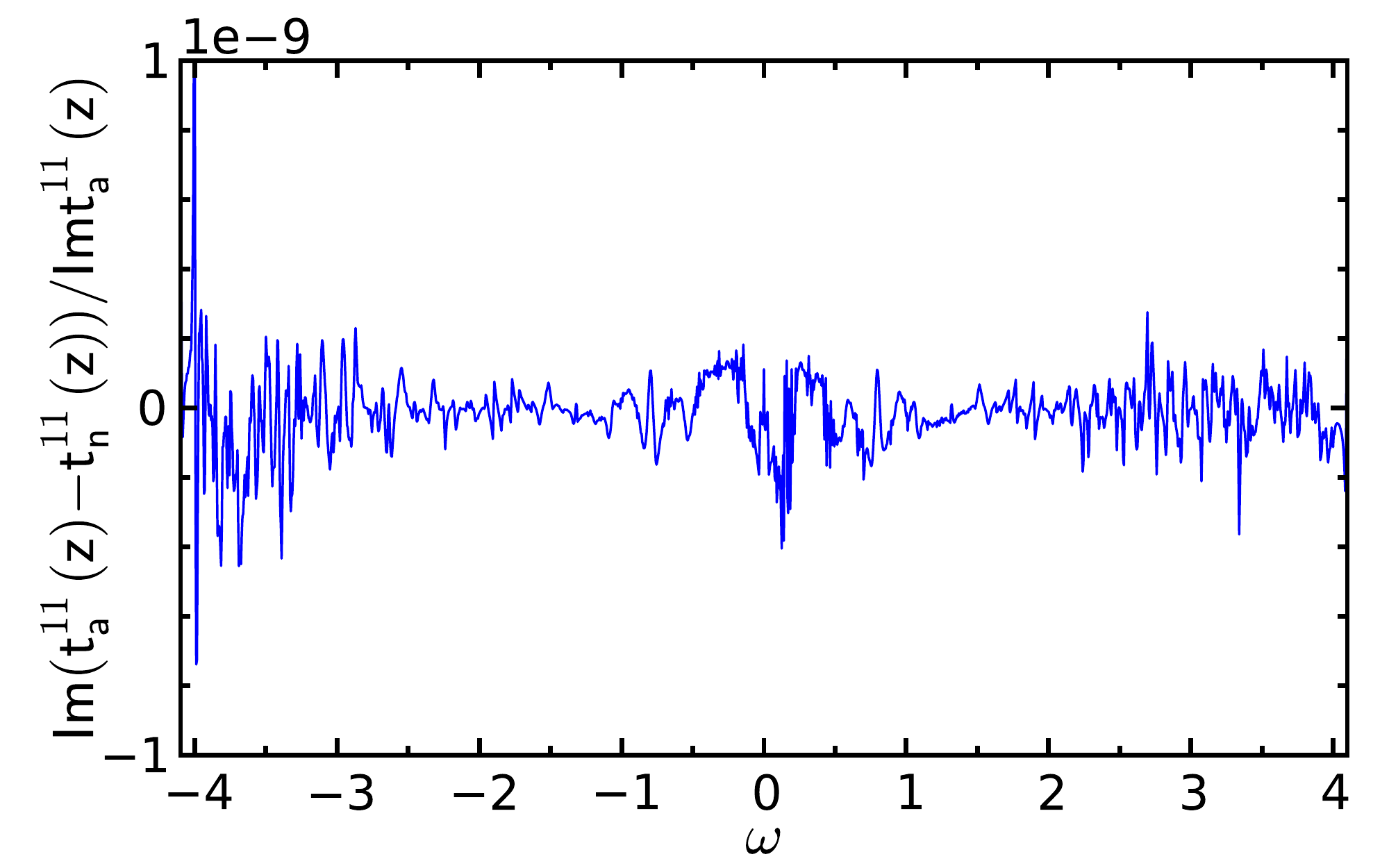}
\caption{\label{fig9}(color online) The relative difference between the
 imaginary part of the analytical ($a$) and numerical ($n$) resolvents
 $t^{11}(z)$ for the finite $N\times N$ cluster with $N=64$, exchanges
 $J=J_0=1$, spins $S=S'=1$, field $h=2$, and $z=\omega+i\eta$ with $\eta=0.005$.
 For these parameters, the uniform canting angle is $\theta=0.2527$ and the
 impurity canting angle is $\theta_i=1.122$.}
\end{figure}

The analytical solution of the artificial problem can also be used to check the
numerical solution beyond the discussion in the main text. To this end,
rather than performing the integration in (\ref{eq:5a}) exactly, the
momentum sum is carried out numerically on a finite cluster of the same size as the one
used in the numerical procedure. The resulting two $T$-matrices can
then be compared. A typical case is shown in Fig.~\ref{fig9}, where we depict
the {\it relative difference} of the diagonal elements $t^{11}(z)$ of the
resolvents obtained from each of the two approaches, versus frequency for a finite
$N\times N$ cluster with $N=64$, exchanges $J=J_0=1$, spins $S=S'=1$, and field
$h=2$. For this plot, the imaginary broadening in $z=\omega+i\eta$ has been set
to $\eta=0.005$. For this system size and the choice of parameters,
such value of $\eta$ is small enough, so that {\em each} individual
delta-function in the spectrum, corresponding to {\em every} single eigenenergy, is
resolved. Therefore, this figure demonstrates an agreement between the two
approaches not only to within the precision of the linear algebra routines that are used
for the numerical Bogolyubov transformation, but also on a level of resolution down to
each individual eigenstate.

\begin{thebibliography}{33}
\bibitem{weaware} After completion of our work, we became aware of the study
\cite{Wollny12}, reporting similar findings but for different types of host spin systems
and impurities.
\end{thebibliography}
\end{document}